\documentclass[12pt]{article}

\usepackage{amssymb}
\usepackage{amsmath}
\usepackage{amscd}
\usepackage{latexsym}
\usepackage{graphicx}

\usepackage{cite}

\newcommand{\be}{\begin{equation}}
\newcommand{\ee}{\end{equation}}
\newcommand{\Dlt}{\Delta}

\newcommand{\bt}{\beta}

\newcommand{\al}{\alpha}
\newcommand{\ra}{\rightarrow}

\newcommand{\gm}{\gamma}

\newcommand{\Gm}{\Gamma}

\newcommand{\lbd}{\lambda}

\begin{document}

\begin{center}

{\Large{\bf Additive self-similar approximants} \\ [5mm]

S. Gluzman and V.I. Yukalov }  \\ [3mm]

{\it Bogolubov Laboratory of Theoretical Physics, \\
Joint Institute for Nuclear Research, Dubna 141980, Russia} 

\end{center}

\vskip 5cm

\begin{abstract}

A novel type of approximants is introduced, being based on the ideas of
self-similar approximation theory. The method is illustrated by the 
examples possessing the structure typical of many problems in applied 
mathematics. Good numerical convergence is demonstrated for the cases 
that can be compared with exact solutions, when these are available. 
The method is shown to be not less and as a rule essentially more 
accurate than that of Pad\'e approximants. Comparison with other 
approximation methods is also given. 

\end{abstract}  

\vskip 3cm

{\parindent=0pt
{\bf Keywords}: Self-similar approximation theory, Additive approximants,
                Numerical convergence

\vskip 2cm

{\it Corresponding author}: V.I. Yukalov, E-mail: yukalov@theor.jinr.ru

}

\newpage

\section{Introduction}

Suppose we are looking for the solution of a complicated equation that
cannot be solved exactly, and the sole thing we are able to accomplish 
is to find asymptotic expansions near the boundary of the solution 
domain. Then the problem arises of reconstructing the sought function
for its whole domain from the limited knowledge of only its asymptotic 
expansions. The most often used method for this purpose is that of 
Pad\'{e} approximants \cite{Baker_1}. Although being useful in many 
cases, this method has a number of known deficiencies that have been
repeatedly discussed in literature (see, e.g., 
\cite{Baker_1,Saff_18,Simon_19,Gluzman_2,Gluzman_3}). Another very
efficient method is based on self-similar approximation theory 
\cite{Yukalov_4,Yukalov_5,Yukalov_6} resulting in several types of 
self-similar approximants \cite{Gluzman_7,Yukalov_8,Gluzman_9}.

In the present paper, we introduce a novel type of self-similar 
approximants and illustrate their effectiveness by examples whose
mathematical structure is typical of many calculational problems in
a variety of applications, e.g., in mathematical chemistry, statistical
physics, nonlinear phenomena, and field theory. We demonstrate that 
the new approximants, that we name {\it additive self-similar 
approximants}, to distinguish them from multiplicative factor 
approximants \cite{Gluzman_7,Yukalov_8,Gluzman_9}, enjoy sufficiently 
fast numerical convergence and provide good accuracy of approximations.  
Their accuracy is not worse, and usually much better, than that of 
Pad\'{e} approximants.

\section{Construction of additive approximants}

Let us be looking for a solution that is a real function $f(x)$ of 
a real variable $x$. In general, the function domain can be arbitrary. 
For concreteness, we consider here the interval $0 \leq x < \infty$.
This does not mitigate the generality of the consideration, since by 
a change of variables it is practically always possible to reduce
a given interval to the ray $[0, \infty)$. 

Suppose that the sought function is defined by complicated equations
that allow us to find only its asymptotic expansion near one of the 
domain boundaries, say, for asymptotically small $x$, where  
\be
\label{1}
f(x) \simeq f_k(x) \qquad ( x \ra 0 ) \;   ,
\ee
with the $k$-th order finite series
\be
\label{2}
 f_k(x) = \sum_{n=0}^k a_n x^n \;  .
\ee
Or the large-variable expansion can be available, such that
\be
\label{3}
 f(x) \simeq f^{(p)}(x) \qquad ( x \ra \infty ) \;  ,
\ee
with the finite series
\be
\label{4}
f^{(p)}(x) =  \sum_{n=1}^p b_n x^{\bt_n} \;   .
\ee
The powers in the above series are arranged in the ascending order:
\be
\label{5}
\bt_n > \bt_{n+1} \qquad ( n =  1,2, \ldots , p-1 ) \;   .
\ee
The standard situation corresponds to the uniform power decrease 
with the constant difference
\be
\label{6}
\Dlt\bt \equiv  \bt_n - \bt_{n+1} \qquad ( n =  1,2, \ldots  ) \;  .
\ee

For the problem of interpolation between the small-variable expansion
(\ref{2}) and large-variable expansion (\ref{4}), we would need the 
values of the coefficients $b_n$. However, the most interesting and 
most complicated problem is that of the extrapolation of the 
small-variable expansion (\ref{2}) into the large-variable limit, 
when the coefficients $b_n$ are not known, although the powers
$\beta_n$ can be available. In the present paper, we shall pay the
main attention exactly to this problem of extrapolation, with unknown
coefficients $b_n$. In that procedure, when the small-variable 
expansion is obeyed by construction, but the large-variable 
coefficients are not known, the error of an appoximation tends to 
zero, as $x \ra 0$, while, vice versa, the error increases when 
the variable tends to infinity, reaching a maximal value in the 
limit $x \ra \infty$. Therefore, in the problem of extrapolation, the 
accuracy of the procedure as a whole is defined by the large-variable 
limit, that is, by the accuracy of the amplitude
\be
\label{7}
 B \equiv \lim_{x\ra\infty} x^{-\bt_1} f(x) = b_1  
\ee
that has to be compared with the large-variable limits of the studied
approximations.

The procedure of extrapolating the small-variable series to the whole
range of the variable $x \in [0, \infty]$, by employing the self-similar
approximation theory has been described in full mathematical details
in our previous papers 
\cite{Yukalov_4,Yukalov_5,Yukalov_6,Gluzman_7,Yukalov_8,Gluzman_9}.
Therefore, here we omit the description of mathematical techniques, 
only stressing the main steps of the procedure, which results in the 
novel type of approximants.

First, we subject the variable $x$ to the affine transformation
$$
x \ra A ( 1 + \lbd x ) \;   ,
$$
consisting of a scaling and shift. This transforms the terms of 
series (\ref{2}) as
$$
 a_n x^n \ra A_n ( 1 + \lbd x)^n \;  ,
$$
where $A_n = A a_n$. Then the self-similar transformation of 
series (\ref{2}) is just the affine transformation of its terms,
which yields
\be
\label{8}
 f_k^*(x) = \sum_i A_i ( 1 + \lbd x)^{n_i} \;   .
\ee
The powers of the first $k$ terms of this series correspond to the 
powers of series (\ref{4}),
\be
\label{9}
 n_i = \bt_i \qquad ( i = 1,2, \ldots, k ) \;  ,
\ee
while all coefficients $A_i$ can be found by the accuracy-through-order
procedure, expanding form (\ref{8}) in powers of $x$ and equating 
to expansion (\ref{2}). Expression (\ref{8}) is the {\it additive
approximant}, which is named for distinguishing it from the 
multiplicative {\it factor approximants} considered earlier
\cite{Gluzman_7,Yukalov_8,Gluzman_9}.  

It is clear that in the large-variable limit, approximant (\ref{8})
will reproduce the terms with the powers of series (\ref{4}). However,
except the terms with the correct powers $\beta_i$, there appear the 
terms with the powers $\beta_i - 1$. There can exist two situations.
It may be that the powers $\beta_i - 1$ do not pertain to the set
of the powers $\{ \beta_i \}$. Then the terms with the incorrect 
powers should be canceled by including in approximant (\ref{8}) 
correcting terms (counter-terms) with the powers
\be
\label{10}  
  n_j = \gm_j \equiv \bt_j -1 \qquad  ( j = 1,2, \ldots, q ) \; ,
\ee
where
\be
\label{11}
\bt_{k+1} < \gm_j < \bt_1 \;    ,
\ee
and the coefficients $C_i$ are defined by the cancellation of the 
terms with incorrect powers in the large-variable expansion. In that
way, the general form of the additive approximant is
\be
\label{12}
 f_{k,q}^*(x) = \sum_{i=1}^k A_i ( 1 + \lbd x)^{\bt_i} +
\sum_{j=1}^q C_j ( 1 + \lbd x)^{\gm_j}  .
\ee

The other possibility is when the powers $\beta_i - 1$ turn out to 
be the members of the set $\{ \beta_i \}$, that is, the set 
$\{ \beta_i \}$ is invariant under the transformation
\be
\label{13}
 \bt_i -1 = \bt_j \;  .
\ee
In that case, no correction terms are needed, and the additive 
approximant is
\be
\label{14}
f_k^*(x) \equiv f_{k,0}^*(x) = \sum_{i=1}^k A_i ( 1 + \lbd x)^{\bt_i} \;   .
\ee
The coefficients $A_i$ can be found by the accuracy-through-order
procedure, comparing the expansion of the additive approximant
with the small-variable expansion, or with the large-variable 
expansion, or using both of them. 

As has been mentioned above, employing the accuracy-through-order
procedure at small variables, the asymptotic expansion at $x\ra 0$
of the additive approximant coincides with the exact asymptotic 
expansion (\ref{2}), while the error of the approximation increases
for growing $x$, reaching the maximal values at $x \ra \infty$. 
It is therefore instructive to compare the exact amplitude (\ref{7})
with the amplitude of the $k$-th approximant 
\be
\label{15}
B_k = \lim_{x\ra\infty} x^{-\bt_1} f_k^*(x) = A_1 \lbd^{\bt_1} \;  .
\ee
 
Of course, not only the leading-order amplitude $B_k$, representing 
the coefficient $b_1$, can be found, but the subleading amplitudes, 
representing other coefficients $b_k$, can also be calculated. However,
our primary interest is not in defining particular coefficients, but
rather to check the accuracy of the whole additive approximant. This 
is why we concentrate our attention on the leading amplitude, given
by (\ref{15}), and characterizing the large-variable limit of the 
additive approximant as a whole.  
   
Defining the coefficients of the additive approximant from the
accuracy-through-order procedure, we confront with the nonuniqueness
of solutions. Thus, when there are no counter-terms, we have $k$
solutions in the $k$-th order. In the case of $q$ counter-terms,
the $k$-th order approximant yields $k+q-1$ solutions. Fortunately,
the appearance of multiple solutions is not a serious obstacle,
because of the following. 

Generally, among the solutions, there can happen real and also 
complex-valued solutions. The latter come in complex conjugate 
pairs, so that their sum is real. It turns out that all real 
solutions and the average sums of the complex conjugate pairs, in each 
order, are very close to each other. We show this in the examples 
below. Then there can be two strategies. Either to consider only 
real solutions, or to take the average sums of all solutions of the 
given order.       
  
\section{Comparison with other methods}

The accuracy of the method of additive approximants will be compared 
with that of other approximation methods. First of all, we consider  
the usual Pad\'{e} approximants $P_{M/N}$, as well as the modified 
$P_{M/N}^\gamma$ Pad\'{e} approximants, suggested by Baker and Gammel
\cite{Baker_10}, where
$$
(M - N) \gm = \bt_1 \;   .
$$
   
It is also important to compare the additive approximants with other
variants of self-similar approximants. As has been mentioned above, 
additive approximants are distinguished from multiplicative factor 
approximants \cite{Gluzman_11,Yukalov_12}. The latter are obtained
by reducing the small-variable asymptotic series 
\be
\label{16}
f_k(x) = f_0(x) \left ( 1 + \sum_{n=1}^k a_n x^n \right )
\ee
to a multiplicative form, accomplishing affine transformations of 
the variable in the multiplicative factors and realizing self-similar 
renormalization yielding the factor approximant
\be
\label{17}
  f_k^*(x) = f_0(x) \prod_{i=1}^{N_k} ( 1 + A_i x)^{n_i} \; ,
\ee
where
\begin{eqnarray}
\nonumber
N_k = \left \{ \begin{array}{ll}
k/2 \; , ~ & ~  k = 2,4,\ldots \\
(k+1)/2 \; , ~ & ~ k = 3,5,\ldots \; ,
\end{array} \right.
\end{eqnarray}
with the powers $n_i$ and coefficients $A_i$ being uniquely defined 
by the accuracy-through-order procedure. 

The zero-order term is usually of the form
$$
 f_0(x) = A x^\al \;  .
$$
Then the large-variable limit of the factor approximant gives the
power
\be
\label{18}
\bt_1 = \al + \sum_{i=1}^{N_k} n_i
\ee
and the amplitude
\be
\label{19}
 B_k = A \prod_{i=1}^{N_k} A_i^{n_i} \;  .
\ee

Another variant of self-similar approximants is represented by 
root approximants \cite{Gluzman_3,Yukalov_13} having the form
\be
\label{20}
 \frac{f_k^*(x)}{f_0(x)} = \left ( \left ( \ldots
(1 + A_1 x)^{n_1} + A_2 x^2 \right )^{n_2} + \ldots + A_k x^k \right )^{n_k} \;  .
\ee
There can exist several cases. If the large-variable expansion is
known, then all coefficients $A_i$ and powers $\beta_i$ are uniquely
defined by the accuracy-through-order procedure at large $x$, giving
\be
\label{21}
j n_j = j+1 - \bt_{k-j} + \bt_{k-j+1} \qquad ( j = 1, 2, \ldots, k-1 )
\ee
and 
\be
\label{22}
n_k = \frac{\bt_1-\al}{k} \qquad (\al\neq \bt_1) \;   .
\ee
As a rule, the difference
\be
\label{23}
\Dlt\bt \equiv \bt_n - \bt_{n+1} = const
\ee
is constant. Then the powers up to $k-1$ are defined by the 
relation
\be
\label{24}
 j n_j = j+1 - \Dlt \bt   \qquad ( j = 1, 2, \ldots, k-1 ) \;  ,
\ee
with $n_k$ as in equation (22). Note that if here $\Delta \beta = 1$,
then $n_j = 1$ up to $j = k - 1$. When in the large-variable expansion,
the powers $\beta_n$ are known, but the coefficients $b_n$ are given
only for $n = 1,2, \ldots,p < k$, then the coefficients $A_n$ of the
root approximant (\ref{20}) are defined by the accuracy-through-order
procedure at small variables, giving $k-p$ equations and at large
variables, giving $p$ equations. If just the sole large-variable 
power $\beta_1$ is given, then setting
\be
\label{25}
n_j = \frac{j+1}{j} \qquad ( j = 1, 2, \ldots, k-1 ) \;  ,
\ee
all coefficients $A_n$ are defined by the accuracy-through-order
procedure at small variable.   

In the following sections, we illustrate the use of the additive 
approximants by the examples, whose mathematical structure is typical
of many applied problems.

\section{Anharmonic partition integral}

The structure of the integral
\be
\label{26}
Z(g) = \frac{1}{\sqrt{\pi}} 
\int_{-\infty}^{\infty} \exp \left ( - z^2 - gz^4 \right ) \; dz
\ee
is typical for numerous problems in quantum chemistry, field theory, 
statistical mechanics, and condensed-matter physics dealing with the
calculation of partition functions, where $g \in [0, \infty)$ plays 
the role of coupling parameter \cite{Sornette_14}. The integral expansion 
at small $g \ra 0$, yields strongly divergent series, with the $k$-th 
order sums
\be
\label{27}
 Z_k(g) = \sum_{n=0}^k c_n g^n \;  ,
\ee
whose coefficients are 
$$
c_n = \frac{(-1)^n}{\sqrt{\pi}\; n!} \; 
\Gm\left ( 2n + \frac{1}{2} \right ) \;   .
$$
The coefficients $c_n$ quickly grow with increasing $n$ tending to
infinity as $n^n$ for $n \gg 1$, which makes the weak-coupling
expansion strongly divergent. At strong coupling, we have
\be
\label{28}
Z(g) \simeq b_1 g^{-1/4} + b_2 g^{-3/4} + b_3 g^{-5/4}
+ b_4 g^{-7/4} \qquad ( g \ra \infty) \;   ,
\ee
where
$$
b_1 = \frac{1}{2\sqrt{\pi} } \; \Gm\left (\frac{1}{4} \right ) = 1.022765\; , \qquad
b_2 = \frac{1}{8\sqrt{\pi} } \; \Gm\left (-\;\frac{1}{4} \right ) = - 0.345684 \; , 
$$
$$
b_3 = \frac{1}{16\sqrt{\pi} } \; \Gm\left (\frac{1}{4} \right ) = 0.127846\; , \qquad
b_4 = \frac{1}{64\sqrt{\pi} } \; \Gm\left (-\;\frac{1}{4} \right ) = - 0.043211 \; . 
$$

The powers of the strong-coupling expansion,
\be
\label{29}
\bt_n = - \; \frac{2n-1}{4}
\ee
enjoy the uniform difference
$$
\Dlt\bt \equiv \bt_n - \bt_{n+1} = \frac{1}{2} \;   .
$$
The set $ \{ \beta_n \}$ is invariant with respect to transformation
(\ref{13}) because of the property
$$
\bt_n - 1 = \bt_{n+2} \;   .
$$
Hence no correction terms are needed. All coefficients $A_i$ of
the additive approximant (\ref{14}) are obtained from the 
accuracy-through order procedure at weak coupling. The error 
of approximants grows as $g \ra \infty$. Therefore the accuracy of 
the method is defined by the accuracy of the strong-coupling 
amplitude
\be
\label{30}
B_k = \lim_{g\ra\infty} g^{1/4} Z_k^*(g) 
\ee
that has to be compared with the exact value $b_1$.

First, we consider only real-valued solutions for $A_i$. In each odd
order, there is just one real solution. Then for the additive 
approximants (\ref{14}), we have to third order
$$
Z_3^*(g) = A_1 ( 1 + \lbd g)^{-1/4} + A_2 ( 1 + \lbd g)^{-3/4} +  
A_3 ( 1 + \lbd g)^{-5/4} \; ,
$$
where
$$
 A_1 = 1.510761 \; , \qquad A_2 = -0.717990 \; , \qquad
A_3 = 0.207229\; , \qquad \lbd = 7.634834 \;  .
$$
This gives the strong-coupling amplitude
$$
B_3 = 0.908858 \qquad (Z_3^* ) \;   .
$$
To fifth order, 
$$
Z_5^*(g) = A_1 ( 1 + \lbd g)^{-1/4} + A_2 ( 1 + \lbd g)^{-3/4} +  
A_3 ( 1 + \lbd g)^{-5/4} + 
$$
$$
+ A_4 ( 1 + \lbd g)^{-7/4} +  A_5 ( 1 + \lbd g)^{-9/4}\;   ,
$$
with the coefficients
$$
A_1 = 1.808031 \; , \qquad A_2 = - 1.543729 \; , \qquad
A_3 = 1.134917 \; , \qquad A_4 = - 0.492745 \;    ,
$$
$$ 
A_5 = 0.093526 \; , \qquad \lbd = 12.297696  \; .
$$
The strong-coupling amplitude is
$$
B_5 = 0.965495 \qquad (Z_5^* ) \;   .
$$

Continuing the procedure, we obtain in higher orders
$$
B_7 = 0.992107 ~~ ( Z_7^* )  \; , \qquad B_9 = 1.005760 ~~ ( Z_9^* ) \; , 
\qquad B_{11} = 1.01312 ~~ ( Z_{11}^* ) \; , 
$$
$$
B_{13} = 1.01720 ~~ ( Z_{13}^* )  \; , \qquad B_{15} = 1.01952 ~~ ( Z_{15}^* ) \; , 
\qquad B_{17} = 1.02085 ~~ ( Z_{17}^* ) \; , 
$$
$$
B_{19} = 1.02072 ~~ ( Z_{19}^* ) \;   .
$$

Comparing these amplitudes with the exact $B = 1.02277$, we find 
the corresponding errors
$$
11\% \; , ~~~ 6\% \; , ~~~ 3\% \; , ~~~ 2\% \; , ~~~ 0.9\% \; , 
~~~ 0.5\% \; , ~~~ 0.3\% \; , ~~~ 0.2\% \;   .
$$
As is seen, the accuracy improves with increasing order, which demonstrates
good numerical convergence.  

Now let us consider the other way, when, in each order, the average of 
all solutions is taken. In the second order, the approximant is
$$
Z_2^*(g) = A_1 ( 1 + \lbd g)^{-1/4} + A_2 ( 1 + \lbd g)^{-3/4} \; .
$$
The weak-coupling accuracy-through-order procedure for the parameters $A_i$ 
gives two complex-conjugate solutions, whose sum is real. The resulting 
strong-coupling amplitude is
$$
B_2 = 0.858304 \qquad (Z_2^* ) \; .
$$

In the third order, there is one real and two complex-conjugate solutions.
Summing them up yields the strong-coupling amplitude
$$
B_3 = 0.915248 \qquad (Z_3^* ) \; .
$$

The fourth order yields two pairs of complex-conjugate solutions,
resulting in the amplitude
$$
B_4 = 0.956250 \qquad (Z_4^* ) \;   .
$$

In the fifth order, there is one real and two pairs of complex-conjugate 
solutions, whose average sum gives 
$$
B_5 = 0.979861 \qquad (Z_5^* ) \;   .
$$

The sixth order produces three pairs of complex-conjugate solutions,
with the corresponding amplitude
$$
 B_6 = 1.000921 \qquad (Z_6^* ) \;  .
$$
  
In the seventh order, there appear one real and three pairs of 
complex-conjugate solutions, giving the amplitude
$$
B_7 = 1.010621 \qquad (Z_7^* ) \;   .
$$

Comparing the obtained $B_k$ with the exact strong-coupling amplitude
$B = b_1 = 1.02277$, we find the errors 
$$
16\% \; , ~~~~ 11\% \; , ~~~~ 7\% \; , ~~~~ 4\% \; , ~~~~ 2\% \; , ~~~~ 1\%  \;  .
$$

Again, we observe good numerical convergence. The accuracy here is a bit better 
than in the case of taking only real solutions, although not much different,
being of the same order. Thus, the seventh-order real approximant has the error 
of $3 \%$, while the sum of all seventh-order approximants gives the error of $1 \%$.    
But dealing with only real solutions is simpler. 

The accuracy of the additive approximants for the studied problem is much better
than that of other approximants. Because of the incompatibility of the powers in 
the weak-coupling and strong-coupling limits, the standard Pad\'e approximants
are not applicable, but the modified Baker-Gammel approximants $P_{N/(N+1)}^{1/4}$      
have to be used. The modified Pad\'e approximant of $19$-th order (with $N=9$)
has an error of $10 \%$, which is much worse than the error of $0.2 \%$ of the 
additive real approximant. Factor approximants are also less accurate. Thus,
the factor approximant of $9$-th order yields an error of $11 \%$, while the additive
approximant in this order exhibits an error of $2\%$. Root approximants for this 
problem are not defined, resulting in complex solutions.

\section{Quartic anharmonic oscillator}

Another example that plays the role of a touchstone for any novel approximation 
method, since it has the structure typical of many applied problems, is the
quartic anharmonic oscillator described by the Hamiltonian
\be
\label{31}
 \hat H = - \;\frac{1}{2} \; \frac{d^2}{dx^2} + \frac{1}{2} \; x^2 + g x^4 \;  ,
\ee
where $x \in (-\infty, \infty)$ and the anharmonicity strength $g \in [0, \infty)$. 

The ground-state energy is given by the lowest eigenvalue of this Hamiltonian. By 
perturbation theory \cite{Bender_17,Hioe_15} with respect to the parameter $g$, one has
\be
\label{32}
 e_k(g) = \sum_{n=0}^k c_n g^n \;  ,
\ee
with the first several coefficients
$$
c_0 = \frac{1}{2} \; , \qquad c_1 = \frac{3}{4} \; , \qquad
c_2 = -\; \frac{21}{8} \; , \qquad c_3 = \frac{333}{16} \; ,
$$
$$
c_4 = -\; \frac{30885}{128} \; , \qquad c_5 = \frac{916731}{256} \; , \qquad
c_6 = -\; \frac{65518401}{1024} \; , \qquad c_7 = \frac{2723294673}{2048} \;   .
$$
The value of these coefficients quickly increases signifying strong divergence
of the expansion.

In the large anharmonicity limit, the finite series
$$
e^{(p)}(g) = \sum_{n=1}^p b_n g^{\bt_n} 
$$
have fractional powers as in the expansion below:
\be
\label{33}
e(g) \simeq b_1 g^{1/3} +  b_2 g^{-1/3} +  b_3 g^{-1} +  b_4 g^{-5/3} +  
b_5 g^{ -7/3} +  b_6 g^{ -3} + b_7 g^{-11/3} + \ldots \; ,
\ee
where $g \ra \infty$ and the first coefficients are
$$
b_1 = 0.667986 \; , \qquad b_2 = 0.143669 \; , \qquad
b_3 = -0.008628 \; ,   
$$
$$
b_4 = 0.000818 \; , \qquad b_5 = -0.000082 \; , \qquad
b_6 = 0.000008 \; .
$$
    
The general form of the powers in the large $g$ expansion
\be
\label{34}
\bt_n =  1 - \; \frac{2n}{3}
\ee
shows that the difference 
$$
\Dlt\bt \equiv \bt_n - \bt_{n+1} = \frac{2}{3}
$$
is constant. But, contrary to the previous case of the anharmonic partition integral,
these powers are not invariant with respect to transformation (\ref{13}). Really,
since
$$
 \bt_m - \bt_n  = \frac{2}{3} \; ( n - m ) \;  ,
$$  
there are no such integers $m$ and $n$ that would give $1$ in right-hand side of the 
above difference. Therefore, correction terms are required, as is explained in Sec. 2. 

Defining all parameters of the additive approximants from the small $g$ expansion,
we again have the situation, where the error increases with growing $g$. Therefore
again the accuracy of the approximants is defined by that of the large $g$ amplitude
\be
\label{35}
B_k = \lim_{g\ra\infty} g^{-1/3} e_k^*(g) \;   .
\ee
  
In order to illustrate the influence of counter-terms, we, first, consider the
additive approximants $e^*_{k,0}$, with real solutions. For instance, to third order, 
we have
$$
e^*_{3,0} = A_1 ( 1 + \lbd g)^{1/3} +  A_2 ( 1 + \lbd g)^{- 1/3} + 
A_3 ( 1 + \lbd g)^{-1} \; ,
$$
for which the accuracy-through-order procedure gives
$$
A_1 = 0.324485 \; , \qquad A_2 = 0.212357 \; , \qquad
A_3 = -0.036842 \; , \qquad \lbd = 10.105351 \; .
$$
The large $g$ amplitude (\ref{35}) is
$$
B_{3,0} = 0.701528 \qquad ( e_{3,0}^* ) \;   .
$$
Continuing calculations to higher orders, we get the amplitudes
$$ 
B_{5,0} = 0.681609 ~~ ( e_{5,0}^* ) \; , \qquad 
B_{7,0} = 0.675129 ~~ ( e_{7,0}^* ) \; , \qquad
B_{9,0} = 0.672345 ~~ ( e_{9,0}^* ) \; ,
$$
$$
B_{11,0} = 0.670931 ~~ ( e_{11,0}^* ) \; , \qquad 
B_{13,0} = 0.670022 ~~ ( e_{13,0}^* ) \; , \qquad
B_{15,0} = 0.669619 ~~ ( e_{15,0}^* ) \; ,
$$
$$
B_{17,0} = 0.669283 ~~ ( e_{17,0}^* ) \; , \qquad 
B_{19,0} = 0.669041 ~~ ( e_{19,0}^* ) \; , \qquad
B_{21,0} = 0.668765 ~~ ( e_{21,0}^* ) \;  .
$$

Comparing the accuracy of the approximants, with respect to the exact amplitude
$B = b_1 = 0.667986$, we obtain the errors
$$
5\% \; , ~~~ 2\% \; , ~~~ 1\% \; , ~~~ 0.7\% \; , ~~~ 0.4\% \; , 
~~~ 0.3\% \; , ~~~ 0.24\% \; , ~~~ 0.19\% \;  ~~~ 0.16\% \; , ~~~ 0.12\% \;  .
$$
Therefore, numerical convergence is achieved even without counter-terms.

Now, let us take into account the required counter-terms. In the first nontrivial 
order, we have
$$
e^*_{2,1} = A_1 ( 1 + \lbd g)^{1/3} +  A_2 ( 1 + \lbd g)^{- 1/3} + 
C_1 ( 1 + \lbd g)^{-2/3} \;   .
$$
The counter-term, with the power $-2/3$, is introduced for cancelling the incorrect 
terms in the large $g$ expansion, as is explained in Sec. 2. The accuracy-through-order 
procedure displays two real solutions giving the amplitudes $0.699953$ and $0.668733$, 
whose average is  
$$   
B_{2,1} = 0.684343 \qquad ( e_{2,1}^* ) \;    .
$$

The next approximant
$$
e^*_{3,1} = A_1 ( 1 + \lbd g)^{1/3} +  A_2 ( 1 + \lbd g)^{- 1/3} + 
A_3 ( 1 + \lbd g)^{-1} + C_1 ( 1 + \lbd g)^{-2/3}  
$$
possesses one real solution and a complex-conjugate pair of solutions for the 
parameters $A_i$ and $C_1$. The real solution gives the amplitude $0.682509$, while
the average of the conjugate pair yields $0.677471$. The average of all solutions 
results in the amplitude
$$
B_{3,1} = 0.679990 \qquad ( e_{3,1}^* ) \;    .
$$

The higher approximant
$$
e^*_{3,2} = A_1 ( 1 + \lbd g)^{1/3} +  A_2 ( 1 + \lbd g)^{- 1/3} + 
A_3 ( 1 + \lbd g)^{-1} + C_1 ( 1 + \lbd g)^{-2/3}  + C_2 ( 1 + \lbd g)^{-4/3} 
$$
contains two counter-terms. For its parameters, we get four solutions composing
two complex-conjugate pairs. The averages of each pair give the amplitudes
$0.672316$ and $0.675572$, which leads to the average amplitude
$$
B_{3,2} = 0.673944 \qquad ( e_{3,2}^* ) \;    .
$$

Continuing in this way, with averaging over all solutions of the same order, we 
obtain the amplitudes
$$
B_{4,2} = 0.670643 ~~ ( e_{4,2}^* ) \; , \qquad 
B_{4,3} = 0.668888 ~~ ( e_{4,3}^* ) \; , \qquad
B_{5,3} = 0.668109 ~~ ( e_{5,3}^* ) \; .
$$
Summarizing, for the approximants with counter-terms, we get the errors
$$
2.4\% \; , ~~~~ 1.8\% \; , ~~~~ 0.9\% \; , ~~~~ 0.4\% \; , ~~~~ 0.1\% \; , 
~~~~ 0.02\% \;    .
$$
This shows numerical convergence and good accuracy. 

Comparing these results with other approximation methods, we see that again
the standard Pad\'{e} approximants are not applicable, because of the 
incompatibility of powers in the small $g$ and large $g$ expansions. The modified
Pad\'{e} approximants $P_{N/(N+1)}^{-1/3}$ of Baker-Gammel \cite{Baker_10} can be 
employed, although their accuracy is much worse than that of the additive 
approximants. For example, the rather high-order modified Pad\'{e} approximant
of $24$-th order has the accuracy of $4 \%$. Factor approximants are more accurate 
than Pad\'{e} approximants, but less accurate than additive approximants. Thus,
the factor approximant of $9$-th order gives the amplitude $B_9 = 0.704391$, which
has the error of $5 \%$. The accuracy of root approximants is between that of
modified Pad\'{e} and factor approximants.

\section{Electron correlation energy}

It is also instructive to illustrate the efficiency of the method for the problems, 
where not many expansion terms are available, but rather just a few. As an example
of such a problem, let us consider the calculation of correlation energy for electron 
gas. The correlation energy of electron gas is usually expressed in dimensionless units 
as a function of the Seitz radius $r_s$. Then the limit of high density corresponds 
to $r_s \ra 0$, while that of low density, to $r_s \ra \infty$. For one-dimensional 
electron gas of high density \cite{Loos_16}, one has
\be
\label{36}
e(r_s) \simeq - \; \frac{\pi^2}{360} + 0.00845 r_s \qquad ( r_s \ra 0 ) \; .
\ee
And the low-density expansion gives
\be
\label{37}
e(r_s) \simeq  \frac{b_1}{r_s} + \frac{b_2}{r_s^{3/2} }  \qquad 
( r_s \ra \infty ) \; ,
\ee
where
$$
 b_1 = - \left ( \ln \sqrt{2\pi} \; - \; \frac{3}{4} \right ) = - 0.168939  \; , \qquad
b_2 = 0.359933 \; .
$$ 
 
The powers in expansion (\ref{37}) have the general form
\be
\label{38}
 \bt_n = -\; \frac{n+1}{2} \;   .
\ee
The power difference 
\be
\label{39}
\Dlt\bt \equiv \bt_n -\bt_{n+1} = \frac{1}{2}
\ee
is constant. The set of the powers $\{ \beta_j \}$ is invariant under transformation (\ref{13}),
since
\be
\label{40}
 \bt_m -\bt_{n} = \frac{1}{2} \; ( n - m ) \; ,
\ee
which yields $\beta_m - \beta_n = 1$ for $m = n -2$. Hence there is no need for counter-terms. 
     
Here we can construct only the low-order additive approximants whose parameters are defined by 
the asymptotic expansions. For instance
$$
e_2^*(r_s) = A_1 ( 1 +\lbd r_s )^{-1} + A_2 ( 1 +\lbd r_s )^{-3/2}  \;  .
$$
There are two complex-conjugate solutions for the parameters, so that the average of the related
forms $e_2^*$ is real. 

The next approximant is 
$$
e_3^*(r_s) = A_1 ( 1 +\lbd r_s )^{-1} + A_2 ( 1 +\lbd r_s )^{-3/2} + 
A_3 ( 1 +\lbd r_s )^{-2}  \;   .
$$
There exist three solutions, a pair of complex-conjugate and one real. In the latter case, the
parameters are
$$
A_1 = -0.040184 \; , \qquad A_2 = 0.041756 \; , \qquad A_3 = - 0.028987 \; ,
\qquad \lbd= 0.237864 \;   .
$$

The accuracy of the additive approximants can be compared with Monte Carlo numerical data 
\cite{Loos_16} available for the range $0 < r_s < 20$. Thus, the maximal error of the real
additive approximant $e_3^*$ is $7\%$ at the point $r_s \approx 15$. The two-point Pad\'{e} 
approximants yield the errors in the range between $2\%$ and $9\%$ at the approximately same 
point $r_s$. For instance, $P_{1/2}(\sqrt{r_s})$ has an error of $2\%$, while $P_{0/3}(\sqrt{r_s})$ 
gives an error of $9\%$. Factor approximants are rather accurate, with an error for $e_3^*$ 
of about $1\%$. But the root approximants are slightly less accurate, giving in third order 
an error of $8\%$.

\section{Conclusion}

We have introduced a novel type of approximants, whose derivation is based on the ideas of
self-similar approximation theory. These approximants enjoy the same asymptotic expansion
as the exact small-variable asymptotic expansion, and possess correct powers in the 
large-variable expansion. The efficiency of the method is illustrated by the examples possessing 
the structure typical of many problems in applied mathematics. Good numerical convergence is 
demonstrated for the cases that can be compared with exact solutions, when a number of terms in
the small-variable asymptotic expansion are available. It is also shown that the approach is 
applicable to the problems having just a few terms in their asymptotic expansions.

Together with other approximation methods, in the frame of self-similar approximation theory, 
additive approximants provide a very efficient new approach for constructing accurate approximate 
solutions for different complicated problems. 

The accuracy of additive approximants is not less and, as a rule, is essentially higher than that 
of Pad\'{e} approximants, including modified Baker-Gammel Pad\'{e} approximants \cite{Baker_10}.
Note that the accuracy of Pad\'{e} approximants can be strongly improved by combining the method 
of self-similar approximants with that of Pad\'{e} approximants, as is demonstrated in 
Ref. \cite{Gluzman_20}.

\vskip 2mm

{\bf Acknowledgment}

\vskip 1mm

One of the authors (V.I.Y.) acknowledges financial support from RFBR (grant $\# 14-02-00723$)
and useful discussions with E.P. Yukalova.

\end{document}